# An Empirical Model for Volatility of Returns and Option Pricing

Joseph L. McCauley and Gemunu H. Gunaratne
Department of Physics
University of Houston
Houston, Texas 77204

PACs numbers:
89.65.Gh   Economics, business, and financial markets
89.75.Fb   Structures and organization in complex systems

**Abstract**

In a seminal paper in 1973, Black and Scholes argued how expected distributions of stock prices can be used to price options. Their model assumed a directed random motion for the returns and consequently a lognormal distribution of asset prices after a finite time. We point out two problems with their formulation. First, we show that the option valuation is not uniquely determined; in particular ,strategies based on the delta-hedge and CAPM (the Capital Asset Pricing Model) are shown to provide different valuations of an option. Second, asset returns are known not to be Gaussian distributed. Empirically, distributions of returns are seen to be much better approximated by an exponential distribution. This exponential distribution of asset prices can be used to develop a new pricing model for options that is shown to provide valuations that agree very well with those used by traders. We show how the Fokker-Planck formulation of fluctuations (i.e., the dynamics of the distribution) can be modified to provide an exponential distribution for returns. We also show how a singular volatility can be used to go smoothly from exponential to Gaussian returns and thereby illustrate why exponential returns cannot be reached perturbatively starting from Gaussian ones, and explain how the theory of 'stochastic volatility' can be obtained from our model by making a bad approximation. Finally, we show how to calculate put and call prices for a stretched exponential density.

## 1. The CAPM portfolio selection strategy

The Capital Asset Pricing Model (CAPM) is very general: it assumes no particular distribution of returns and is consistent with any distribution with finite first and second moments. Therefore, in this section, we generally assume the empirical distribution of returns but also will apply the model to Gaussian returns (lognormal prices) in part 2 below. The CAPM is not, as is often claimed, an equilibrium model because the distribution of returns is not an equilibrium distribution. We will exhibit the time-dependence of some of the parameters in the model in the familiar lognormal price approximation. Economists and finance theorists (including Sharpe [1] and Black [2]; see also Bodie and Merton [3])) have adopted and propagated the strange notion that random motion of returns defines 'equilibrium', which disagrees with the requirement that in equilibrium no averages of any moment of the distribution can change with time. Random motion in the market is due to trading and the excess demand of unfilled limit orders prevents equilibrium at all or almost all times. Apparently, what many economists mean by 'equilibrium' is more akin to assuming the EMH (efficient market hypothesis), which has nothing to do with vanishing excess demand in the market. The only dynamically consistent definition of equilibrium is vanishing excess demand: if p denote the price of an asset then excess demand ε is defined by dp/dt= ε(p,t) including the case where the right-hand side is drift plus noise, as in stochastic dynamical models of the market. These issues have been discussed in detail in a previous paper [4]. Bodie and Merton [3] claim that vanishing excess demand is necessary for the CAPM, but one sees in part 2 below that no such assumption comes into play during the derivation and would even cause *all* returns to vanish in the model!

The CAPM [5] can be stated in the following way: Let $R_o$ denote the risk-free interest rate,

$$x_k = \ln(p_k(t + \Delta t)/p_k(t))$$

(1)

is the fluctuating return on asset k where $p_k(t)$ is the price of the kth asset at time t. The total return x on the portfolio of n assets relative to the risk free rate is given by

$$x - R_o = \sum_{i=0}^{n} f_i(x_i - R_o)$$

(2)

where $f_k$ is the fraction of the total budget that is bet on asset k. The CAPM minimizes the mean square fluctuation

$$\sigma^2 = \sum_{i,j} f_i f_j \langle (x_i - R_o)(x_j - R_o) \rangle = \sum_{i,j} f_i f_j \sigma_{ij}$$

(3)

subject to the constraints of fixed expected return R,

$$R - R_o = \langle (x - R_o) \rangle = \sum_{i,j} f_i \langle (x_i - R_o) \rangle = \sum_{i,j} f_i (R_i - R_o)$$

(4)

and fixed normalization

$$\sum_{i=0}^{n} f_i = 1$$

(5)

where $\sigma_{ij}$ is the correlation matrix

$$\sigma_{ij} = \langle (x_i - R_o)(x_j - R_o) \rangle$$

(6)

Following Varian, we solve

$$\sum_i \sigma_{ki} f_i = \sigma_{ke} = \sigma_{ee}(R_k - R_o)/\Delta R_e$$

(7)

for the f's where $\Delta R_e = R_e - R_o$ and $R_e$ is the expected return of the 'efficient portfolio', the portfolio constructed from f's that satisfy the condition (7). The expected return on asset k can be written as

$$f_1/f_2 = (\sigma_{12}\Delta R_2 - \sigma_{22}\Delta R_1)/(\sigma_{12}\Delta R_1 - \sigma_{11}\Delta R_2)$$

(8)

where $\sigma_{ee}$ is the mean square fluctuation of the efficient portfolio, $\sigma_{ke}$ is the correlation matrix element between the kth asset and the efficient portfolio, and $\beta\Delta R_e$ is the risk premium for asset k.

For many assets n in a well-diversified portfolio, studying the largest eigenvalue of the correlation matrix $\sigma$ seems to show that that eigenvalue represents the market as a whole, and that clusters of eigenvalues represent sectors of the market like transportation, paper, etc. [6]. However, in formulating and deriving the CAPM above, nothing is assumed either about diversification or how to choose a winning portfolio (the strategies of agents like Buffet and Lynch have not been mathematized and apparently do not depend on the CAPM notion of diversification and risk minimization), only how to try to minimize the fluctuations in any arbitrarily-chosen portfolio of n assets, which portfolio may or may not be well-diversified relative to the market as a whole, and which may well consist of a basket of losers. Negative x represents a short position, positive x a long position. Large beta implies both greater risk and larger expected return. Without larger expected return a trader will not likely place a bet to take on more risk. Negative returns R can and do occur systematically in market downturns, and in other bad bets.

We define a liquid market as one where an agent can reverse his trade over a very short time interval $\Delta t$ with only very small transaction costs and net losses, as in the stock market on the scale of seconds during normal trading. A market crash is by definition a liquidity drought where limit orders placed for selling overwhelmingly dominate limit orders placed for buying. Large deviations in the theory of Gaussian returns (lognormal price distribution) are by far too unlikely to match the empirical data on crashes and bubbles.

In what follows we consider a portfolio of 2 assets, e.g. a bond (asset #1) and the corresponding European call option (asset # 2). For two assets the solution for the CAPM portfolio can be written in the form needed in part 2 below,

$$\Delta R_k = \frac{\sigma_{ke}}{\sigma_{ee}} \Delta R_e = \beta_k \Delta R_e$$

(9)

Actually there are 3 assets in this model because a fraction $x_o$ can be invested in a risk free asset, or may be borrowed in which case
$x_o < 0$.

## 2. Black-Scholes theory of option pricing

Let p denote the price of asset #1, the bond, e.g., and w(p,t) the price of a corresponding European call option. In this section, in order to discuss the original Black-Scholes derivation [7], we follow Osborne [8] and assume that asset prices are distributed lognormally, with stochastic equation

$$\Delta p = rp\Delta t + \sigma_1 p \varepsilon \Delta t^{1/2}$$

(10)

where ε is a Gaussian random variable with null mean and unit variance, and $\sigma_1$ is assumed constant. The right hand side of this equation is simply the excess demand ε times Δt, dp/dt=ε(p,t), as is emphasized in [4]. This term does not vanish, either in the market or in the CAPM, nor (due to limit orders) does the total excess demand of the market. *There is no equilibrium, either in the market or in the model.* The stochastic equation for the price change of the option is then

$$\Delta w = \frac{\partial w}{\partial t}\Delta t + \frac{\partial w}{\partial p}\Delta p + \frac{1}{2}\frac{\partial^2 w}{\partial p^2}\Delta p^2 = w_t \Delta t + w' \Delta p + \frac{1}{2} w'' \Delta p^2$$

(11)

to O(Δt). In equations (10) and (11) the initial data p(t) and w(p,t) are deterministic at the first instant t while the changes Δp and Δw as well as p(t+Δt) and w(p+Δp,t+Δt) are random due to ε. The object is to derive a deterministic diffusive equation of motion for the option price w(p,t) by taking expectation values. Two methods are presented in the original Black-Scholes paper [7].

### 2.1. The Delta Hedge Strategy

The standard idea is to try to construct a risk-neutral hedge [7,9]. This is done by choosing the portfolio so that the leading random terms in the two stochastic equations (10) and (11) cancel each other. The delta hedge, which is defined by a portfolio with value

$$\Pi = -w + w' p$$

(12)

does this. The return on the portfolio, with w' fixed during a very short trading interval Δt, is then given by

$$R_\Pi = \frac{\Delta \Pi}{\Pi \Delta t} = (-\Delta w + w' \Delta p)/\Pi \Delta t = (-w_t \Delta t - w'' \Delta p^2 / 2)/\Pi \Delta t$$
(13)

or

$$R_\Pi = (-w_t \Delta t - w'' \sigma^2 \varepsilon^2 / 2)/\Pi \Delta t$$
(13b)

First, we follow the standard treatment and set $\varepsilon^2 = 1$ in the last term in the return (13b), replacing a randomly fluctuating term in the stochastic equation (11) by a deterministic one [7,9,10] to obtain

$$\Delta w \approx w_t \Delta t + w' \Delta p + \frac{1}{2} w'' \sigma_1 p^2 \Delta t$$

(14)

This, as we show below, leads to an error when compared with the correct calculation of the mean square fluctuation in the option return. Using (14) first, the return (13) is deterministic and the standard arbitrage argument sets the portfolio return equal to the risk free interest rate,

$$R_\Pi = \frac{\Delta \Pi}{\Pi \Delta t} \approx (-w_t - w'' \sigma p^2 / 2)/\Pi = R_o$$

(15)

This yields the so-called risk-neutral hedge and the Black-Scholes equation

$$R_o w = w_t + R_o p w' + \sigma_1 p^2 w'' / 2$$

(16)

for the option price w(p,t). Much has been made of the fact that agents' preferences do not enter into this equation [11].

In the texts on finance theory and stochastic differential equations $\varepsilon^2$ in $\Delta p^2$ (13) is set equal to unity 'with probability one' whenever $\varepsilon$ is a Gaussian process. It is assumed that self-averaging occurs in the limit where Δt goes to zero. In reality, tics in the market are on the order of seconds so that no such self-averaging is possible for small Δt. Aside from market realities of discrete tics, there is also a mathematical reason why one should not assume a priori that Ito calculus applies to finance: as a $0^{th}$ order approximation logarithmic returns $x=\ln(p(t)/p(0))$ are distributed normally. This means that the random variable $\varepsilon(\Delta t)$ in

$$\frac{\Delta x - R\Delta t}{\sigma \Delta t^{1/2}} = \varepsilon(\Delta t)$$

(10b)

is a Gaussian independent variable for all $\Delta t$ regardless of the value of $p(t)/p(0)$. The starting point of finance text arguments assumes something quite different, namely that

$$\frac{\Delta p/p - R'\Delta t}{\sigma \Delta t^{1/2}} = \varepsilon(\Delta t)$$

(10c)

is a Gaussian independent variable for all t independent of the size of $\Delta p/p$, which is not true and is inconsistent with (10b). Usually, (10c) is assumed to be Gaussian and (10b) is then derived via Ito calculus, but this is an inconsistent procedure because the distribution of $\Delta p/p$ is nonGaussian and is quite complicated, for arbitrary $\Delta p/p$. Stated clearly, $\varepsilon(\Delta t)$ in (2) is not a Gaussian independent variable for arbitrary values of $\Delta p/p$. For an example of such conflicting equations see (10.7) and (11.1) in Hull [9].

Aside from replacing the fluctuating term by a deterministic one there remains a small mistake: traders do not use the risk free interest rate, but instead add 2-3% onto $R_o$ in the B-S equation. The reason for this is the profit motive: a sensible trader would not likely go to the trouble to construct a complicated portfolio that must be updated continually (dynamic hedging) just to get the risk free interest rate. He would likely construct this portfolio only if he expects to make a profit over and above the risk free rate $R_o$. Otherwise he should simply buy a CD or use the money market. We call this the profit motive, which is neglected in the usual arbitrage argument. In general, arguments based on neo-classical economic theory eliminate the possibility of profit, but traders exist purely for profit and producers exist largely for profit [12].

If one goes on to calculate the mean square fluctuation in the return using the approximation employed above in (14), then one obtains

$$\sigma_{22} = (\frac{pw'}{w})^2 \sigma_{11}$$

(17)

and

$$\langle \Delta r_\Pi^2 \rangle = 0$$

(18)

to $O(\Delta t^2)$ so that the portfolio is predicted to be risk free in the variance to within at least $O(\Delta t)$. We will next show that there is an additional reason to increase the yield on the delta-hedge portfolio. The portfolio is not risk-free to O(Δt) as was previously assumed but fluctuates in value at that level.

We calculate the mean square fluctuation in the portfolio return again, this time treating correctly the fluctuating term in $\varepsilon^2$ approximated above as deterministic. The mean square fluctuation is given by

$$\sigma_\Pi^2 = (\sigma_1^2 p^2 w'' \Delta t / 2)^2 \langle (\varepsilon^2 - 1)^2 \rangle$$

(19)

so that the variance of the portfolio increases like Δt, meaning that the hedge must be rebalanced more often than previously thought. In addition to wanting a profit over and above the risk-free rate of return $R_o$, the trader will increase $R_\Pi$ even further above $R_o$ to compensate himself for the riskiness of the hedge.

Note that in any case the ratio invested is given by

$$f_1 / f_2 = -pw' / w$$

(20)

We will need this result below for comparison with the corresponding CAPM strategy of option pricing, and will see, in contrast with the claim of the original Black-Scholes paper [7], that these two strategies do not and cannot agree with each other, even in the limit where Δt goes to zero. In all that follows we do not approximate the fluctuating term $O(\varepsilon^2)$ in (11) by a deterministic one but calculate correctly (we do not use Ito calculus as presented in finance texts [9,10]).

## 2.2. The CAPM option pricing strategy

In this case the average return on the option is given from (10) by

$$R_2 = \frac{\langle \Delta w \rangle}{w \Delta t} = \frac{w_t}{w} + \frac{pw'}{w} R_1 + \frac{1}{2} \sigma_1^2 p^2 \frac{w''}{w}$$

(21)

and from CAPM by

$$R_2 = R_o + \beta_2 \Delta R_e$$

(22)

whereas the average return on the stock is given from CAPM by

$$R_1 = R_o + \beta_1 \Delta R_e$$

(23)

According to Black and Scholes [7], we should be able to prove that

$$\beta_2 = \frac{pw'}{w}\beta_1$$

(24)

Were this the case then, combining (21), (22) and (23), we would get a cancellation of the two beta terms in (25) below:

$$R_2 = R_o + \beta_2 \Delta R_e = \frac{w_t}{w} + \frac{pw'}{w}R_1 + \frac{1}{2}\sigma_1^2 p^2 \frac{w''}{w} = \frac{w_t}{w} + \frac{pw'}{w}R_o + \frac{pw'}{w}\beta_1 \Delta R_e + \frac{1}{2}\sigma_1^2 p^2 \frac{w''}{w}$$

(25)

leaving us with only the risk-free rate of return and the risk-neutral options pricing pde (16). To see that (24) is a wrong assumption, simply use it to calculate the fractions $f_2$ and $f_1$ invested in the CAPM portfolio. One then finds that $f_1$ is finite while $f_2 = 0$, so that the fraction invested in the option must be zero! This is in stark contrast with (20) above obtained for the delta-hedge strategy.

Equation (24) is in fact impossible to derive without making a serious error. Within the context of CAPM there is no reason to use (24) in (25). When we calculate the fluctuating return on the option we obtain

$$r_2 - R_o = R_2 - R_o + \frac{pw'}{w}(r_1 - R_1) + \frac{w' \sigma_1^2 p^2}{2w}(\varepsilon^2 - 1)$$

(26)

so that the average return calculation merely yields

$$\langle r_2 - R_o \rangle = R_2 - R_o$$

(27)

which is true but does not lead to (24), in contrast with the claim in Black-Scholes [7]. To go further, calculate the ratio invested $f_2/f_1$ by our hypothetical CAPM risk-minimizing agent. Here, we need the correlation matrix only to leading order in $\Delta t$:

$$\sigma_{11} \approx \sigma_1^2 / \Delta t$$

(28)

$$\sigma_{12} \approx \frac{pw'}{w}\sigma_{11}$$

(29)

and

$$\sigma_{22} \approx (\frac{pw'}{w})^2 \sigma_{11}$$

(30)

It is then easy to show, to leading order in Δt, that

$$f_1 \propto (\beta_1 pw'/w - \beta_2)pw'/w$$

(31)

and

$$f_2 \propto (\beta_2 - \beta_1 pw'/w)$$

(32)

so that it is impossible that (24) could be satisfied! Note that the ratio $f_1/f_2$ is the same as for the delta-hedge.

That CAPM is not an equilibrium model is exhibited explicitly by the time dependence of the neglected terms in (28-30). When one keeps the higher order terms in Δt then $f_1$. $f_2$, $\sigma_{ij}$ and $\sigma_2$ all become time dependent. It would be a serious mistake to try to think of an arbitrary time t dynamically as a point of equilibrium: the self-contradiction in the economists' notion of 'temporary price equilibria' [13] has been exhibited elsewhere [4].

The CAPM simply does not predict either the same option pricing equation as does the delta-hedge. Furthermore, if traders actually use the delta-hedge in option pricing then this means that agents do not trade in a way that minimizes the mean square fluctuation ala CAPM. The CAPM and the delta-hedge do not try to reduce risk in exactly the same way. In the delta-hedge the main fluctuating terms are removed directly from the portfolio return, thereby lowering the expected return, whereas in CAPM nothing is subtracted from the return in forming the portfolio and the idea there is not only diversification but also increased expected return through increased risk. This is illustrated explicitly by the fact that the expected return on the hedged portfolio is not the risk-free return, but is instead proportional to the factor set equal to zero by Black and Scholes, shown above as equation (24):

$$\Delta R = \frac{\beta_1 pw'/w - \beta_2}{pw'/w - 1} \Delta R_e$$

(33)

Note also that the expected return ΔR in excess of the risk-free rate depends on time, not only through the term pw'/w, but also through the terms of higher order neglected in (28-30), even if the β's were t-independent (but we know that they are not). Note also from (33) that beta for the CAPM hedge is given by

$$\beta_{CAPM} = \frac{\beta_1 pw'/w - \beta_2}{pw'/w - 1}$$

(34)

The notion of increased expected return via increased risk is not present in the delta-hedge strategy, which tries to eliminate risk and to minimize return. We see now that the way that options are priced is strategy-dependent, which may be closer to the notion that psychology plays a role in trading. The CAPM option pricing equation depends on the expected returns for both stock and option,

$$R_2 w = w_t + pw' R_1 + \frac{1}{2}\sigma_1^2 p^2 w''$$

(35)

and so differs from the original Black-Scholes equation (16) of the delta-hedge strategy There is no such thing as a universal option pricing equation independent of the chosen strategy, even if that strategy is reflected in this era by the market. Economics is not like physics (non-thinking nature), but depends on human behavior and expectations [14,15].

Perhaps, if one were clever enough, one could invent expected utilities that would be maximized by either the delta-hedge strategy or the CAPM strategy, but what would be the point? The idea of utility is superfluous in the construction of either model.
.
We turn now to the empirical distribution, and to theory that goes beyond Black-Scholes. What follows is based on the empirical distribution and a stochastic theory describing it instead of the idea of fudging the Black-Scholes equation via the financial engineering trick of using implied or stochastic volatility. With the right distribution, the empirical one, volatility should anyway be correctly described.

### 3. An empirical model for option pricing

### 3.1 Introduction

We observe that returns of bonds and foreign exchange are approximately exponentially distributed. The exponential distribution is used to develop a pricing formula that agrees with market values. It follows that out of the money options are not overpriced as is assumed on the basis of lognormal pricing theory, and that traders are smarter than theorists.

We begin by asking which variable should be used to describe the variation of the underlying asset price p. Suppose p changed from $p(t)$ to $p(t+\Delta t)=p+\Delta p$ in the time interval from t to $t+\Delta t$. Price p can of course be measured in different units (e.g, ticks, Euros, Yen or Dollars), but we want our equation to be independent of the units of measure, a point that has been ignored in many other recent data analyses. E.g., the variable $\Delta p$ is additive but is units-dependent. The obvious way to achieve independence of units is to study $\Delta p/p$, but this variable is not additive. This is a serious setback for a theoretical analysis. A variable that is both additive and units-independent is $x=\ln(p(t)/p(0))$, in agreement with Osborne [8] who reasoned from Fechner's Law and was the first econophysicist. In this notation $\Delta x = \ln(p(t+\Delta t)/p(t))$. We will study x in this section but will also use $\Delta x$ in part 4 below. According to Dacorogna et al [16], correct tail exponents for very large deviations (so-called 'extreme values') for the empirical distribution cannot be obtained unless one studies logarithmic returns x.

The basic assumption of the theory of option pricing is that the variables $x(t)$ and $x(t')$ are statistically independent. This assumption allows the use of the Central Limit Theorem, leading to a Gaussian distribution of returns (lognormal prices, as first proposed by Osborne [8]) for long enough time increments. The assumption of statistical independence is a necessary evil; it is obviously untrue but is essential for the development of a complete theory. We will present some consequences of relaxing the assumption in studying the empirical distribution.

We begin the next section with one assumption, and then from the historical data for US Bonds and for two currencies we show that the distribution of returns x is in fact much closer to exponential than to Gaussian. This has consequences for theory, as we show in part 4: a theoretical approach that begins in $0^{th}$ order with exponential returns has a far greater chance of success in describing the data than does one that begins, in $0^{th}$ order, with Gaussian returns. After describing some useful features of the exponential distribution we present a model for option pricing that uses two undetermined parameters. We show how the parameters can be estimated and discuss some important consequences of the new model. We finally compare the theoretically predicted option prices with actual market prices.

Throughout this section the option prices given by formulae refer to European options. When the need arises to determine the value of an American option we can use the quadratic approximation to evaluate the early exercise premium.

## 3.2 Black-Scholes theory

The primary assumption of the Black-Scholes theory is that the successive x's are distributed independently and identically. The lognormal price distribution then follows from the central limit theorem, at long enough times.

$$x(t) = \ln(p(t)/p(0)) \to N(\mu \Delta t, \sigma^2 \Delta t)$$
(36)

where N is the Gaussian distribution with mean

$$\mu \Delta t = (R - \sigma^2/2)\Delta t$$

(37)

σ is the variance, and Δt=T-t is the time to expiration (T is the strike time). In the integrals below we have x=ln(p/p(t)) where p is the integration variable and p(t)=$p_o$ is the price at time t obtained from diffusion backward in time via the Black-Scholes -equation.

Given the above assumptions European options are priced by by their average values at expiration, giving the call price as

$$C(K, p_o, \Delta t) = \langle (p-K)\theta(p-K) \rangle = e^{-R\Delta t} \int_{\ln n(K/p_o)}^{\infty} (p-K)P(x, \Delta t)dx$$

(38)

and the put price as

$$P(K, p_o, \Delta t) = \langle (K-p)\theta(K-p) \rangle = e^{-R\Delta t} \int_{0}^{\ln(K/p_o)} (K-p)P(x, \Delta t)dx$$

(39)

The time evolution of the price of an option is due in part to the decrease in time to expiration T, and partly due to the variation of $\sigma^2 \Delta t$. Under the assumption of independent events $\sigma^2$ is constant, in disagreement with the known behavior of the volatility. That the average volatility behaves as

$$\langle (x - \langle x \rangle)^2 \rangle = \sigma^2 \Delta t^H$$

(40)

With H=O(1) is known to be approximately correct empirically for $\Delta t$>10 minutes in trading [17], but this is not a test for statistical independence of the x's.

It is known empirically that options far from the money generally trade at a higher price than in Black-Scholes theory [9]. This is academically attributed to overpricing. The deviation is taken into account by considering the implied volatility as a function of strike price K. Further frequent 'runs' in price suggest the importance of correlations between successive price increments, thus once more undermining the assumed independence of the x's. Long runs in one direction are of course very unlikely events if the successive increments were independent.

### 3.3 Exponential distribution

The objections raised above lead us to analyse the actual distribution of returns x, and to see if any conclusion can be drawn about their distribution. The frequencies of returns for US Bonds and some currencies are shown in figures 1, 2, and 3. It is clear from the histogram, at least for short times $\Delta t$, that the logarithm of the price increment p(t)/p(0), x, is distributed very close to an exponential that is generally skew. We describe some properties of the new distribution here and deduce it's consequences for the pricing of options in part 3.4.

It is important to observe that the deviations from linearity seen in figures 1-3 are due to the strong correlations between successive price increments. If their existence is assumed then the frequent occurrence of 'runs' is immediately explained. We will present a dynamical theory for the exponential distribution in part 4 below where correlations in returns and volatility are neglected.

The tails of the exponential distribution fall off much more slowly than those of normal distributions, so that large fluctuations in returns is much more likely. Consequently, the price of out of the money options will be larger than that given by the Black-Scholes theory.

Suppose that the price of an asset moves from p(0) to p(t) in time t. Then we assume that the variable $x = \ln(p(t)/p(0))$ is distributed with density

$$P(x,t) = A e^{\gamma(x-\delta)} \quad x < \delta$$

$$P(x,t) = A e^{-\nu(x-\delta)} \quad x > \delta$$

(41)

Where δ, γ and ν are the parameters that define the distribution. The normalization coefficient is given by

$$A = \frac{\gamma \nu}{\gamma + \nu}$$

(42)

Consequently, the density of the variable y=p(t)/p(0) has fat tails,

$$f(y,t) = Ae^{-\gamma\delta} y^{\gamma-1}, y < e^{\delta}$$

$$f(y,t) = Ae^{-\gamma\delta} y^{-\nu-1}, y > e$$
(43)

The parameters γ and ν are discussed both empirically and semi-theoretically in part 3.4.

Typically, a large a mount of data is needed to get a definitive form for the histograms as in figures 1-3. With smaller amounts of data it is generally impossible to guess the correct form of the distribution. Before proceeding let us describe a scheme to deduce that the distribution is exponential as opposed to normal or truncated symmetric Levy. The method is basically a comparison of mean and standard deviation for different regions of the distribution. Define

$$\langle x \rangle_+ = \int_\delta^\infty xP(x,t)dx = \delta + \frac{1}{\nu}$$

(44)

to be the mean of the distribution for x>δ

$$\langle x \rangle_- = \int_{-\infty}^{\delta} xP(x,t)dx = \delta - \frac{1}{\gamma}$$
(45)

as the mean for that part with x<δ. The mean of the entire distribution is

$$\langle x \rangle = \delta + \frac{(\gamma - \nu)}{\gamma \nu}$$
(46)

The analogous expressions for the mean square s are

$$\langle x^2 \rangle_+ = \frac{2}{\nu^2} + 2\frac{\delta}{\nu} + \delta^2$$
(47)

$$\langle x^2 \rangle_- = \frac{2}{\gamma^2} - 2\frac{\delta}{\gamma} + \delta^2$$

and

(48)

Hence the variances for the distinct regions are

$$\sigma_+^2 = \frac{1}{\nu^2} : \sigma_-^2 = \frac{1}{\gamma^2}$$

(49)

And for the whole

$$\sigma^2 = \sigma_+^2 + \sigma_-^2 = \frac{1}{\nu^2} + \frac{1}{\gamma^2}$$

(50)

For $\Delta t = .5 - 4$ hours $\gamma$ and $\nu$ are on the order of 500. With

$$\langle x \rangle_\pm = R_\pm \Delta t$$
(51)

we then have

$$\delta \approx R_\pm \Delta t$$

(52)

where

$$(R_+ - R_-)\Delta t = \langle x \rangle_+ - \langle x \rangle_- = \frac{1}{\nu} - \frac{1}{\gamma} \approx 0$$

(53)

for the time scales $\Delta t$ of data analysed here. Hence the quantities $\gamma$ and $\nu$ can be calculated from a given set of data. The average of x is generally small and should not be used for comparisons, but one can check if the relationships between the quantities are valid for the given distribution. Their validity will give us confidence in the assumed exponential distribution. The two relationships that can be checked are $\sigma^2 = \sigma_+^2 + \sigma_-^2$ and $\sigma_+ + \sigma_- = x_+ + x_-$. Our histograms do not include extreme values of x where P decays like a power of x [16], and we also do not discuss results from intraday trading.

Given that the average of the volatility obeys

$$\sigma^2 = \langle (x - \langle x \rangle)^2 \rangle = c\Delta t^{2H}$$

(54)

Where H=O(1/2) and c=constant, we see that the fat tail exponents in (43) decrease with time,

$$\gamma = 1/b' \Delta t^{H/2}$$
$$\nu = 1/b\Delta t^{H/2}$$

(55)

Where b and b' are constants. In our data analysis we find that the exponential distribution spreads consistent with 2H on the order of unity, although whether 2H ≈ 1, ..9, or 1.1, we cannot determine at this stage. We will next see that the divergence of γ and ν as Δt vanishes is necessary for correct option pricing near the strike time. In addition, only the choice H=1/2 is consistent with our assumption of statistical independence. For H≠1/2 in (54) one has fractional Brownian motion with persistence or antipersistence [18]. We therefore assume that H=1/2 in all that follows.

### 3.4 Option pricing

Our starting point for option pricing is the assumption that the call prices are given by averaging over the final option price max(p-K,0) with the exponential
(56)

$$C(K, p_o, \Delta t) = \langle (p-K)\theta(p-K) \rangle = e^{-R\Delta t} \int_{\ln(K/p_o)}^{\infty} (p-K)P(x,\Delta t)dx$$

distribution, and puts by

$$P(K, p_o, \Delta t) = \langle (K-p)\theta(K-p) \rangle = e^{-R\Delta t} \int_{-\infty}^{\ln(K/p_o)} (K-p)P(x,\Delta t)dx$$

(57)

where P(x, Δt) is the exponential density (41) of returns. Here, $p_o$ is the price at time t and the strike occurs at time T, where Δt = T-t. In order to determine δ empirically here we use the traders' idea that the average stock price increases at the cost of carry c,

$$\langle p/p_o \rangle = \langle e^x \rangle = e^{c\Delta t}$$
(58)

which yields

$$\delta = c\Delta t + \ln((\nu-1)(\gamma+1)/\nu\gamma)$$
(59)

If we now take R=R$_o$ then including c is equivalent to assuming that the hedge is not risk-neutral, that the trader adds a few percentage points onto the risk-free rate of interest in his hedge.

For the exponential density of returns we find that the call price of a strike K at time T is given for $x_K=\ln(K/p_o) < \delta$ by

$$C(K,p_o,\Delta t)e^{R_o\Delta t} = p_o e^{\delta}\frac{\gamma\nu}{\gamma+\nu} - K + \frac{KAe^{-\gamma\delta}}{\gamma(\gamma+1)}(\frac{K}{p_o})^{\gamma}$$
(60)

where $p_o$ is the underlying futures and A and $\delta$ are given by (42) and (59). For $x_K > \delta$ the call price is given by

$$C(K,p_o,\Delta t)e^{R_o\Delta t} = \frac{KAe^{\nu\delta}}{\nu(\nu-1)}(\frac{K}{p_o})^{-\nu}$$
(61)

Observe that, unlike in the standard theory, these expressions and their derivatives can be calculated explicitly. The corresponding put prices are given by

$$P(K,p_o,\Delta t)e^{R_o\Delta t} = \frac{KAe^{-\gamma\delta}}{\gamma(\gamma+1)}(\frac{K}{p_o})^{\gamma}$$
(62)

for $x_K < \delta$ and by

$$P(K,p_o,\Delta t)e^{R_o\Delta t} = K - p_o e^{\delta}\frac{\gamma\nu}{(\gamma+1)(\nu-1)} + \frac{KAe^{\nu\delta}}{\nu(\nu-1)}(\frac{K}{p_o})^{-\nu}$$
(63)

for $x_K > \delta$.

Note that the initial condition C = max(p-K,0) = (p-K)θ(p-K) is reproduced by these solutions as $\gamma$ and $\nu$ go to infinity. Using this limit the density of returns (41), we see that P(x,t) peaks sharply at x=$\delta$ and is approximately zero elsewhere. A standard largest term approximation (via Watson's lemma [19]) in (56) yields

$$Ce^{R\Delta t} \approx (p_o e^{\delta} - K)\theta(p_o e^{\delta} - K)\int_K^{\delta} p_-(x,t)dx + (p_o e^{\delta} - K)\theta(p_o e^{\delta} - K)\int_{\delta}^{\infty} p_+(x,t)dx$$
$$= (p_o e^{\delta} - K)\theta(p_o e^{\delta} - K) \approx (p_o - K)\theta(p_o - K)$$
(64)

as $\delta$ vanishes. For $x_K > \delta$ we get C=0 whereas for $x_K < \delta$ we retrieve C=($p_o$-K), as required. Therefore, our pricing model recovers the initial condition for calls at strike time T, and likewise for the puts (62) and (63).

All that remains empirically is to estimate the two parameters $\gamma$ and $\nu$ from data (we do not attempt to determine b, b' and H empirically here). We outline a scheme that is useful when the parameters vary in time. We assume that the options close to the money are priced correctly, i.e., according to the correct frequency of occurrence. Then by using a least squares fit we can determine the parameters $\gamma$ and $\nu$. We typically use six option prices to determine the parameters, and find the rms deviation is generally very small; i.e., at least for the options close to the money, the expressions (60) - (63) give consistent results. Note that when fitting, we use the call prices for the strikes above the future and put prices for those below. These are the most often traded options, and hence are more likely to be traded at the 'correct' price.

Table 1 shows a comparison of the results with actual prices. The option prices shown are for the contract US89U whose expiration day was 18 August 1989 (the date at which this analysis was performed). The second column shows the end-of-day prices for options (C and P denote calls and puts respectively), on 3 May 1989 with 107 days to expiration. Column C gives the equivalent annualized implied volatilities assuming Black-Scholes theory. The values of $\gamma$ and $\nu$ are estimated to be 10.96 and 16.76 using prices of three options on either side of the futures price 89.92. The rms deviation for the fractional difference is 0.0027, suggesting a good fit for six points. Column 4 shows the prices of options predicted by equations (60-63). (We have taken into account the fact that options trade in discrete tics, and have chosen the tic price by the number larger than the actual price. We have added a price of 0.5 tics as the transaction cost). The last column gives the actual implied volatilities from the Black-Scholes formulae. Columns 2 and 4, as well as columns 3 and 5, are almost identical, confirming that the options are indeed priced according to the proper frequency of occurrence in the entire range. Figure 4 compares the implied volatilities with those determined from equations (60-3). Note that in all of the above calculations we have used the quadratic approximation [9] to evaluate the early exercise option.

The model above contains a flaw, the option prices can blow up and go negative at extremely large times $\Delta t$ where $\nu \leq 1$ (the integrals (56-7) diverge for $\nu=1$). But since the annual value of $\nu$ is roughly 10, the order of magnitude of the time required for divergence is about 100 years. This is irrelevant for trading. More explicitly, $\nu = 540$ for I hour, 180 for a day (assuming 9 trading hours/day) and 10 for a year, so that $b=1/540 hour^{1/2}$.

We now exhibit the dynamics of the exponential distribution, which, due to our assumption of statistical independence of returns, requires $H=1/2$. The dynamics of exponential returns leads inescapably to a dynamic theory of volatility, in contrast with the standard theory.

## 4. Dynamics of volatility of returns and option pricing

We use the delta-hedge strategy and extend option pricing dynamics to include exponential and other possible distributions of returns that are far from Gaussian but which reduce to Gaussian (Black-Scholes Theory) if a certain parameter is set equal to zero in a formula for volatility displayed in part 5. An important point is that our result for exponentially-distributed returns cannot be reached via perturbations starting with Gaussian returns (lognormal prices) because the perturbation is highly singular. Our generalization is also consistent with the assumptions of the CAPM, where no particular dynamics need be assumed. Our analysis is motivated by part 3 above and by the possibility that drift and diffusion coefficients for the stochastic equation below might be extracted from empirical data, but here we infer the diffusion coefficient from the empirical distribution of part 3 combined with the standard requirement that average volatility should show Brownian-like behavior. So far, no one has exhibited drift and diffusion coefficients empirically for returns although interesting results have been obtained for small price differences [20,21]. For the delta hedge we do not need the drift coefficient, only the diffusion term. In our case we infer the diffusion coefficient by asking which coefficient is required to obtain the exponential distribution, with *average* mean square fluctuation $\sigma^2 \sim \Delta t$, from a Fokker-Planck-type equation (for Fokker-Planck equations, see [22]. We also find that the scaling exponents of fat tails [16,17,23] are not constant but decrease with time. For studies of distributions of returns, see Dacorogna [16].

## 4.1 Black-Scholes theory rewritten for returns

We begin by rewriting the Black-Scholes theory as a pde for returns. With $x(t) = \ln(p(t)/p(0))$ the stochastic equation is

$$\Delta x = (r - \sigma_1^2 \varepsilon^2 / 2)\Delta t + \sigma_1 \varepsilon \Delta t^{1/2}$$

(65)

where $\Delta x = \ln(p(t+\Delta t)/p(t))$ with $\varepsilon$ a Gaussian random variable and p is the underlying asset price. The Fokker-Planck equation for the distribution $P(x,t)$ of asset returns is

$$P_t = -((r - \sigma_1^2/2)P)' + \frac{\sigma_1^2}{2} P''$$

(66)

Let $u(x,t) = w(p,t)$ denote the option price. Transforming (16) from p to x yields the Black-Scholes equation in the simple form

$$Ru = u_t + (R - \frac{\sigma_1^2}{2})u' + \frac{\sigma_1^2}{2}u''$$

(67)

where (following traders instead of theorists) R must be large enough to induce the trader to take the risk of constructing the delta hedge and updating it frequently. With

$$u = e^{Rt}v$$
(68)

we get

$$v_t = -(R - \sigma_1^2/2)v' - \frac{\sigma_1^2}{2}v''$$

(69)

and the forward-time initial condition at t=T is just

$$v(x,T) = \max(e^x p_o - K, 0)$$
(70)

where K is the strike price at time T. With $\Delta t = T-t$, the call solution backward in time is

$$C(K,p_o,\Delta t)e^{R\Delta t} = \int_{-\infty}^{\infty} G(x,\Delta t)v(x,T)dx = \int_{-\infty}^{\infty} G(x,\Delta t)(p-K)\theta(p-K)dx = \langle(p-K)\theta(p-K)\rangle$$

(71)

where the Green function G(x,Δt) of (71) is the Gaussian distribution

$$G(x,\Delta t) = N(R - \sigma_1^2/2, \sigma_1^2)$$

(72)

Inserting (72) into (71), European call prices are given by

$$C(p_o,K,\Delta t) = e^{-R\Delta t}\int_{\ln K/p_o}^{\infty} G(x,\Delta t)(p-K)dx$$

(73)

where

$$p = p_o e^x$$

(74)

and the corresponding put prices are

$$P(p_o,K,\Delta t) = e^{R\Delta t}\int_{-\infty}^{\ln K/p_o} G(x,\Delta t)(K-p)dx$$

(75)

With a simple transformation of variables in the integrand these equations lead to the standard results of Black-Scholes theory [9].

The above formulation is what we now generalize to include other return distributions, including a certain fat tailed one, the exponential distribution.

### 4.2 Volatility

In this section we assume a diffusion coefficient/volatility of the form

$$D = D(x,t,\alpha)$$

(76)

Imagine a model where with parameter α at a certain limit we would retrieve the Black-Scholes theory of Gaussian returns, whereby

$$D = b^2$$

(76b)

is a constant, but where another value of $\alpha$ is required for the theory of exponential returns, which is described below by a purely singular anomalous diffusion coefficient with nontrivial volatility. The theory of stochastic volatility for H=1/2 can be formulated as follows. Beginning with a stochastic equation

$$\Delta x = R\Delta t + (D(x,t))^{1/2} \varepsilon \Delta t^{1/2}$$
(77)

with $\varepsilon$ generally a nonGaussian independently distributed random variable. Averaging over $\varepsilon$ is equivalent to averaging over all possible jumps $\Delta x$ starting with initial condition x(t). We can average over $\varepsilon$ (but not over initial data x) to obtain the fluctuating (in x) volatility

$$\langle \Delta x^2 \rangle = D(x,t)\Delta t$$
(78)

The volatility is just the diffusion coefficient, which fluctuates with returns x. Using the equation describing local conservation of probability

$$P_t = -RP' + \frac{1}{2}(DP)''$$
(79)

to calculate the probability density P(x,t), the average of the volatility follows from averaging over x with the density P(x,t),

$$\sigma^2 = \langle\langle \Delta x^2 \rangle\rangle = (\int D(x,t)P(x,t)dx)\Delta t = \langle D(x,t) \rangle \Delta t$$
(80)

where the average over x of D must be independent of $\Delta t$ in order for (77) to make sense (i.e., H=1/2). This is a general way to formulate the theory of returns and volatility for H=1/2. We now apply to show how the exponential returns of part 3 can be obtained from a singular volatility, or vice-versa.

### 4.3 Dynamics of the exponential distribution

In our stochastic equation for asset returns (stock, bond or foreign exchange)

$$\Delta x = R_+\Delta t + (D_+(x,t))^{1/2} \varepsilon \Delta t^{1/2}$$
(81)

$\varepsilon$ is an exponentially independent and identically distributed random variable with null mean and unit variance, so that $\Delta x$ is the random variable and x is the initial condition at time t. The plus subscript denotes the region $x>\delta$ and $R_+$ is the expected return for that part of the distribution. We take the expected return to be piecewise constant. This makes sense because x is itself the fluctuating return. The probability conservation equation for this region is

$$P_t = -R_+P' + \frac{1}{2}(D_+P)''$$

(82)

In order to describe the log-exponential distribution of prices, or exponential returns, we assume next that

$$D_+(x,t) = b^2 \nu(x-\delta)$$

(83)

when $x > \delta$ and

$$D_-(x,t) = b'^2 \gamma(\delta - x)$$
(84)

when $x < \delta$. To understand this qualitatively, first assume that the distribution is "softer" when the deviation from the mean is larger. The simplest form is $D(x,t) \sim (x-\delta)$. Now, looking at the exponential distribution, it is clear that this quantity needs to be scaled by $\nu$, leading us to (83). Replacing x by $x_K = \ln(K/p_o)$, note also that our diffusion coefficient (83), (84) yields an approximation to 'volatility smile'. Finally, no other assumption leads from the probability conservation equation to the exponential distribution of returns (41) with $\gamma$ and $\nu$ given by (55) with $H=1/2$.

We can start with our volatility, make a mistake, and end up with the theory of stochastic volatility. Begin with (41) and (83) and write

$$P(x,t) \propto e^{-\nu(x-\delta)} = e^{-(\nu(x-\delta))^2/D/b^2} = e^{-(x-\delta))^2/D\Delta t}$$
(41b)

Now, make the error of assuming that volatility D is distributed differently than x and you get 'stochastic volatility theory'. In the simplest case x and D are uncorrelated [9], which is completely unrealistic. In our model x and D are perfectly correlated but with 'returns lag' $\delta$. In general, x and D are perfectly correlated in any Smoluchoski model because D is a function of (x,t). After all, 'volatility' is nothing other than big swings in returns over short time intervals.

Substituting the diffusion coefficient (83) for D in the Smoluchowski equation (82) yields

$$\frac{A_t}{A} + \nu\delta_t + \nu_t\delta - \nu_t x = \nu R_+ + -b\nu^2 + b\nu^3(x-\delta)/2$$

(85)

so that equating coefficients of x yields the equation

$$-\nu_t = b^2\nu^3/2$$
(86)

and the result

$$v = 1/b(\Delta t)^{1/2}$$
(87)

Correspondingly, for the region x<δ we use an average return R. and diffusion coefficient

$$D_-(x,t) = b'^2 \gamma (\delta - x)$$
(88)

To obtain

$$\gamma = 1/b' (\Delta t)^{1/2}$$
(89)

Now, averaging over initial conditions x in (83) and using the condition

$$\langle x \rangle_+ = \delta + \frac{1}{v}$$
(90)

we obtain the average volatility for x>δ

$$\langle\langle \Delta x^2 \rangle\rangle_+ = \langle D_+(x,t) \rangle \Delta t = b^2 v (\langle x \rangle_+ - \delta) \Delta t = b^2 \Delta t$$
(91a)

and likewise

$$\langle\langle \Delta x^2 \rangle\rangle_- = \langle D_-(x,t) \rangle \Delta t = b'^2 \gamma (\delta - \langle x \rangle_-) \Delta t = b'^2 \Delta t$$
(91b)

in agreement with (49) of part 3.

In contrast with the theory of Gaussian returns, the volatility (83-4) is singular, vanishes for small returns but diverges for very large returns. Volatility, like returns, is exponentially distributed but yields Brownian-like mean square fluctuation (91a,b) on the average.

Going further, if we use (84) and the analogous equations for x<δ to calculate δ then we obtain

$$\delta = R_+ \Delta t - \frac{1}{v}$$
(92)

and

$$\delta = R_- \Delta t + \frac{1}{\gamma}$$
(93)

so that (92-3) agree with the predictions (44) and (45) for the average return calculated directly from the exponential distribution (41), where we identify

$$\langle x \rangle_+ = R_+ \Delta t$$
(94)

and

$$\langle x \rangle_- = R_- \Delta t$$
(95)

Note also that

$$(R_+ - R_-)\Delta t = \frac{1}{\gamma} - \frac{1}{\nu}$$
(96)

Because of this condition only three of the four parameters ($R_+$, $R_-$, b, b') are independent and are free to be fixed by the empirical data.

It is the solution for $\delta$ in the two regions $x < \delta$ and $x > \delta$ that forces the absence of an additive constant term is (83-4), so that the diffusion coefficient is purely singular for exponential returns. We turn now to option pricing, which is determined largely by the volatility $D(x,t)$.

### 4.4 Option pricing with the exponential distribution

We now break further with Black, Scholes, and Merton by formulating the theory of hedging from the start in terms of returns. The reason for this was pointed out in part 3 above: $\Delta p$ is an unsuitable variable empirically because it is not units free, and $\Delta p/p$ is an unsuitable variable both theoretically and empirically because it is not additive. For all distributions of returns we therefore start with the delta hedge in the form

$$\Pi = -u + u'$$
(97)

where the portfolio return is then

$$\frac{\Delta \Pi}{\Pi \Delta t} = \frac{-\Delta u + u' \Delta x}{(-u + u')\Delta t}$$
(98)

Here, it is u'(x,t) that is held constant by the trader during a time interval $\Delta t$, which differs from holding w'(p,t) constant. The average return R for the hedge portfolio is given by

$$\frac{\langle \Delta \Pi \rangle}{\Pi \Delta t} = R$$
(99)

The equation of motion for the average option price u(x,t) is then

$$R u = u_t + R u' + \frac{D_\pm}{2} u''$$

(100)

where plus and minus subscripts denote the two regions of interest for the exponential distribution (41).

To solve (100), we first write (with Δt = T-t)

$$u = e^{-R_\pm t} v$$

(101)

to obtain

$$0 = v_t + r R_\pm v' + \frac{D_\pm}{2} v''$$

(102)

which still contains the time dependence of δ, γ and ν, as we will show.

Since the option prices must be written as initial value problems

$$C(K, p_O, \Delta t) = e^{-R\Delta t} \int_{x_K}^{\infty} (p_o e^x - K) v(x, \Delta t) dx$$

(103)
and

$$P(K, p_O, \Delta t) = e^{-r\Delta t} \int_{0}^{x_K} (p_o e^x - K) v(x, \Delta t) dx$$

(104)

there are several facts that guide us in solving (102). First, the solution v must depend on x-δ with δ given by (92-3), just as it appears in the diffusion coefficient (83-4). Second, we must have the same normalization factor A as in (41) so that the integral of v over all x is normalized to unity in order to recover the initial condition C=max(P-K,0) at time T for the call, and P=max(K-p,0) for the put, via Watson's lemma [19]. These considerations yield an exact solution of the form

$$v(x, \Delta t) = A e^{-\nu(x - \delta - 2\Delta r_+ / b^2 \nu^2 - \ln \nu / \nu)}$$

(105)

for x>δ, where Δr$_+$ = (R$_+$ - R), and

$$v(x, \Delta t) = Ae^{\gamma(x-\delta+2\Delta r_-/b'^2 \gamma^2 - 2\ln\gamma/\gamma)}$$
(106)

with $\Delta r_- = R_- - R$ for $x<\delta$. Using (105) and (106) to evaluate (103) and (104) yields our option pricing predictions, where 3 of the 4 parameters (b, b', R+, R-) must be determined by the data (R is typically a few percentage points higher than the risk-free rate of interest). There are fewer free parameters in the standard theory, but the standard theory disagrees very badly with market data.

The results for option pricing are qualitatively the same as those obtained in part 3, with minor quantitative differences. For example, there is no need to insert the cost of carry extra, and δ is determined by the equations (92-3) where R+ and R- must be estimated from historic data or, more likely, from traders' expectations. In what follows let

$$\delta_\pm = \delta + \alpha_\pm$$
(107)

Where α is the deviation from δ given in the exponents in (105) and (106). The exact solution for calls is then given by

$$C(K, p_o, \Delta t)e^{R\Delta t} = A\left(\frac{p_o e^{\delta+v\alpha_+}}{v-1} - \frac{K}{v}e^{v\alpha_+}\right) + \frac{Ap_o}{\gamma+1}e^{\delta-\gamma\alpha_-} - \frac{KA}{\gamma}e^{-\gamma\alpha_-}$$
(108)

for $x_K<\delta$ and by

$$Ce^{R\Delta t} = Ae^{-vx_K}\left(\frac{p_o e^{v(\delta+\alpha_+)}}{v-1} - \frac{KA}{v}\right)$$
(109)

for $x_K>\delta$. Puts are given by

$$P(K, p_o, \Delta t)e^{R\Delta t} = \left(\frac{KA}{v} - \frac{p_o Ae^\delta}{v-1}\right)\left(e^{-v(x_K-\delta_+)} - e^{-v(\delta-\delta_+)}\right) + Ae^{\gamma(x_K-\delta_-)}\left(\frac{K}{\gamma} - \frac{p_o}{\gamma+1}\right)$$
(110)

for $x_K<\delta$ and

$$P(K, p_o, \Delta t)e^{R\Delta t} = \frac{KA}{v(v-1)}e^{-v(x_K-\delta_+)}$$
(111)

for $x_K>\delta$. For α = 0, and δ given by (59), we retrieve the predictions (60-63) of part 3.4: the results are qualitatively the same and are quantitatively very close to one another.

## 5.1 Volatility, market crashes, and scaling exponents

Our discontinuous diffusion coefficient/fluctuating volatility (83-4) may be regarded as an approximation to a smooth one with a minimum at x=δ, analogous to volatility smile. This corresponds to the possibility that the exponential distribution may be inaccurate at very small returns x, where the peak in the histograms in figures (1-3) is approximated by a discontinuous slope. A constant diffusion coefficient (76b) might be required in order to describe small returns correctly at large times Δt.

Note that even if we were to try to begin perturbatively with a volatility of the form

$$D_+(x,t,\alpha) \approx b^2 + \delta\alpha f(x,t)$$
(76c)

where δα is small, and for very small returns x≈Δp/p, then the stochastic equation (81) for returns could not be reached perturbatively in Δp by starting with Black-Scholes theory formulated in terms of the variable p, because the anomalous diffusion term in (83-4) goes like x=ln(p(t)/p(0)), requiring the resummation of infinitely many terms in perturbation theory (i.e., perturbation theory would lead to a mess). We conclude therefore that methods based on stochastic volatility, ARCH and GARCH [9] cannot lead to a correct option pricing theory because the chosen variable in those models is price p, not return x. In part 5.2 below we show how the exponential density must be generalized to lead more or less smoothly to a Gaussian density and constant volatility. But what about scaling?

The exponential distribution, rewritten in terms of the variable y=p/p(0), has fat tails with time-dependent tail exponents γ and ν. These tail exponents become smaller as Δt increases. The probability density

$$\tilde{P}(y,t) = P(\ln y, t)/y$$

(112)

shows self-affine scaling in y (i.e., in p) with the same two t-dependent exponents γ and ν. However, trying to rewrite the dynamics in terms of p or Δp rather than x leads to excessively complicated equations, in contrast with the simplicity of the theory above written in terms of x. From our standpoint the scaling itself is not particularly useful or important in applications like option pricing, nor is it helpful in understanding the underlying dynamics. In fact, concentrating on scaling would have sidetracked us from looking in the right direction for the solution. Also, one cannot discover correct fat tail exponents for extreme values of x without studying the distribution of x [16]. Using returns, it is also easy to do VAR [24] and obtain risk estimates considerably different from those obtained using the theory of Gaussian returns.

Note that while we need at least the first two moments of asset returns for option pricing, one usually settles for the average or expected option price u(x,t)=w(p,t). We could also write down a diffusion equation for the option price probability density, but fluctuations in option prices are typically ignored in practice except during market crashes and other times of high volatility. Perhaps a description of fluctuations in option prices will also become of interest in the future, at least for the estimation of option prices during large-σ events like market crashes, where σ is the variance (54) for logarithmic returns x. We end with a simple description of option price fluctuations.

From

$$\Delta u = u_t \Delta t + u' \Delta + u'' \Delta x^2 / 2$$

(113)

we obtain, for x<δ by averaging over ε, that

$$\frac{\langle \Delta u^2 \rangle_-^{1/2}}{u} \approx \left| \frac{u''}{u} \right| b'^2 (\delta - x) \langle \varepsilon^4 \rangle^{1/2} \Delta t^{1/2}$$
(114)

Due to volatility D, fluctuations in option price can dominate the average option price u if the return x is large enough, e.g. when, roughly speaking, x<<0, as during a market crash, where x is large and negative while Δt is relatively small. More accurately, the change in volatility for x<δ is given by

$$\Delta D_- = -b'^2 \nu \Delta x$$

(115)

A large event is described by Δx large in magnitude over a short time interval Δt. The description of a sequence of large events ('runs') is a sequence of large changes in volatility.

Large volatility makes accurate option pricing more difficult or impossible. In contrast, were we to use instead the average volatility (91a,b) then we would obtain (for x<δ) the nonvolatile Black-Scholes-like prediction

$$\frac{\langle\langle \Delta w^2 \rangle\rangle_-^{1/2}}{w} \approx \left| \frac{u''}{u} \right| b'^2 \Delta t^{1/2}$$
(114b)

which tells us nothing about the accuracy of the predicted average price w during times of high volatility.

### 5.2 Interpolating singular volatility

We can interpolate from exponential to Gaussian returns with the following volatility,

$$D_+(x,t) = b^2 (\nu(x-\delta))^{2-\alpha}, x > \delta$$

$$D_-(x,t) = b'^2 (\gamma(\delta-x))^{2-\alpha}, x < \delta$$
(116)

where 1≤α≤2 is constant. We do not know which probability density solves the local probability conservation equation to lowest order (79) with this diffusion coefficient, except that it is not a *simple* stretched exponential of the form

$$P(x,t) = Ae^{-(\nu(x-\delta))^\alpha}, x > \delta$$

$$P(x,t) = Ae^{(\gamma(x-\delta))^\alpha}, x < \delta$$

(117)

However, whatever the solution is it interpolates between exponential and Gaussian returns, with one proviso. In order for this claim to make sense we would have to retrieve
(118)

$$\langle D_+ \rangle_+ = b^2 \int_\delta^\infty (\nu(x-\delta))^{2-\alpha} p(x,t) dx = b^2 n$$

where n is independent of Δt, otherwise (116) could lead to fractional Brownian motion, violating our assumption of statistical independence of returns. Equation (118) will hold for any density that has the scaling form (e.g., for x>δ) p((ν(x-δ)$^\alpha$)).

It's important to note that the required diffusion coefficient/volatility is 'singular all the way' as α varies from 2 to 1. As soon as we depart by 'epsilon' from the Gaussian density, setting 2-α=δα with δα>0 small, then the volatility (116) is nonconstant and singular, e.g.

$$D_+ = (\nu(x-\delta))^{\delta\alpha} = e^{\delta\alpha \ln(\nu(x-\delta))} \approx 1 + \delta\alpha \ln(\nu(x-\delta)) + ...$$
(119)

Therefore perturbation theory, starting from a Gaussian density, would require the resummation of an infinite series of logarithmic terms to get the right probability density, which solves the probability conservation equation with (116). In other words, the attempt to describe exponential returns perturbatively starting from a Gaussian would be futile.

## 6. Option pricing via stretched exponentials

Although we do not understand the dynamics of the stretched exponential density (117) we can still use it, in the spirit of part 3.4 above, to price options, if the need should arise empirically. First, using the integration variable

$$z = (\nu(x-\delta))^\alpha$$

(120)

and correspondingly

$$dx = \nu^{-1} z^{1/\alpha - 1} dz$$
(120b)

we can easily evaluate all averages of the form

$$\langle z^n \rangle_+ = A \int_\delta^\infty (\nu(x-\delta))^{n\alpha} e^{-(\nu(x-\delta))^\alpha} dx$$
(121)

for n an integer. Therefore we can reproduce correct versions of (42)-(50) of part 3. For example,

$$A = \frac{\gamma \nu}{\gamma + \nu} \frac{1}{\Gamma(1/\alpha)}$$
(42b)

where $\Gamma(\zeta)$ is the Gamma function, and

$$\langle x \rangle_+ = \delta - \frac{1}{\nu} \frac{\Gamma(2/\alpha)}{\Gamma(1/\alpha)}$$
(44b)

Calculating the mean square fluctuation is equally simple, but without an underlying dynamics we cannot assert a priori that H=1/2 when $1<\alpha<2$, although we suspect that it is true.

Option pricing for $\alpha \neq 1$ leads to integrals that must be evaluated numerically. For example, the price of a call with $x_K > \delta$ is

$$C(K,p_o,\Delta t)e^{R\Delta t} = \frac{A}{\nu}(e^{\nu\delta} p_o \int_{z_K}^\infty e^{\nu^{-1}z^{1/\alpha}} z^{1/\alpha - 1} e^{-z} dz - K\Gamma(1/\alpha, z_K))$$
(122)

where

$$z_K = (\nu(x_K - \delta))^\alpha$$
(123)

and $\Gamma(1/\alpha, z_K)$ is the incomplete Gamma function. The analogues of (60) and (62-63) are equally easy to write down. Retrieving initial data at the strike time follows as before via Watson's lemma.

**Acknowledgement**

One of us (GHG) is grateful to collegues at the TradeLink Corporation during 1990. The other (JMC) is grateful to Chairman Larry Pinsky and the Econophysics Program at the University of Houston Department of Physics for encouragement including financial support, and to Kevin Bassler and George Reiter for very helpful discussions and comments during the Fall Econphysics Seminar at UH. JMC is also grateful to M. Dacorogna, J. Peinke, and L-H Tang for friendly and helpful discussions. We are also grateful to Harry Thomas for a critical comment on Ito's lemma.

## Figure Captions

1. The histogram for the distribution of relative price increments for US Bonds for a period of 600 days. The horizontal axis is the variable $x = \ln(p(t+\Delta t)/p(t))$, and the vertical axis is the logarithm of the frequency of it's occurrence ($\Delta t=4$ hours). The piecewise linearity of the plot implies that the distribution of returns x is exponential.

2. The histogram for the relative price increments of Japanese Yen for a period of 100 days with $\Delta t=1$ hour.

3. The histogram for the relative price increments for the Deutsche Mark for a period of 100 days with $\Delta t=0.5$ hours.

4. The implied volatilities of options compared with those using equations (60-63) (solid line). This plot is made in the spirit of 'financial engineering'. The time evolution of $\gamma$ and $\nu$ is described by (55), and a fine-grained description of volatility is presented in part 4 below.

## Tables

1. Comparison of an actual price distribution of options with the results given by (60-63). See the following text for details. The good agreement of columns 2 and 4, as well as columns 3 and 5, confirms that the options are indeed priced according to the distribution of relative price increments.

| Strike Price and Type | Option Price | Implied Volatility | Computed Option Price | Computed Implied Vol. |
|---|---|---|---|---|
| 76P | 0.047 | 0.150 | 0.031 | 0.139 |
| 78P | 0.063 | 0.136 | 0.047 | 0.129 |
| 80P | 0.110 | 0.128 | 0.093 | 0.128 |
| 82P | 0.172 | 0.116 | 0.172 | 0.117 |
| 84P | 0.313 | 0.109 | 0.297 | 0.108 |
| 86P | 0.594 | 0.104 | 0.594 | 0.104 |
| 88P | 1.078 | 0.100 | 1.078 | 0.100 |
| 90P | 1.852 | 0.095 | 2.859 | 0.096 |
| 92P | 3.000 | 0.093 | 2.984 | 0.093 |
| 94C | 0.469 | 0.093 | 0.469 | 0.093 |
| 96C | 0.219 | 0.094 | 0.219 | 0.094 |
| 98C | 0.109 | 0.098 | 0.109 | 0.098 |
| 100C | 0.047 | 0.100 | 0.063 | 0.104 |
| 102C | 0.016 | 0.098 | 0.031 | 0.106 |
| 104C | 0.016 | 0.109 | 0.016 | 0.109 |

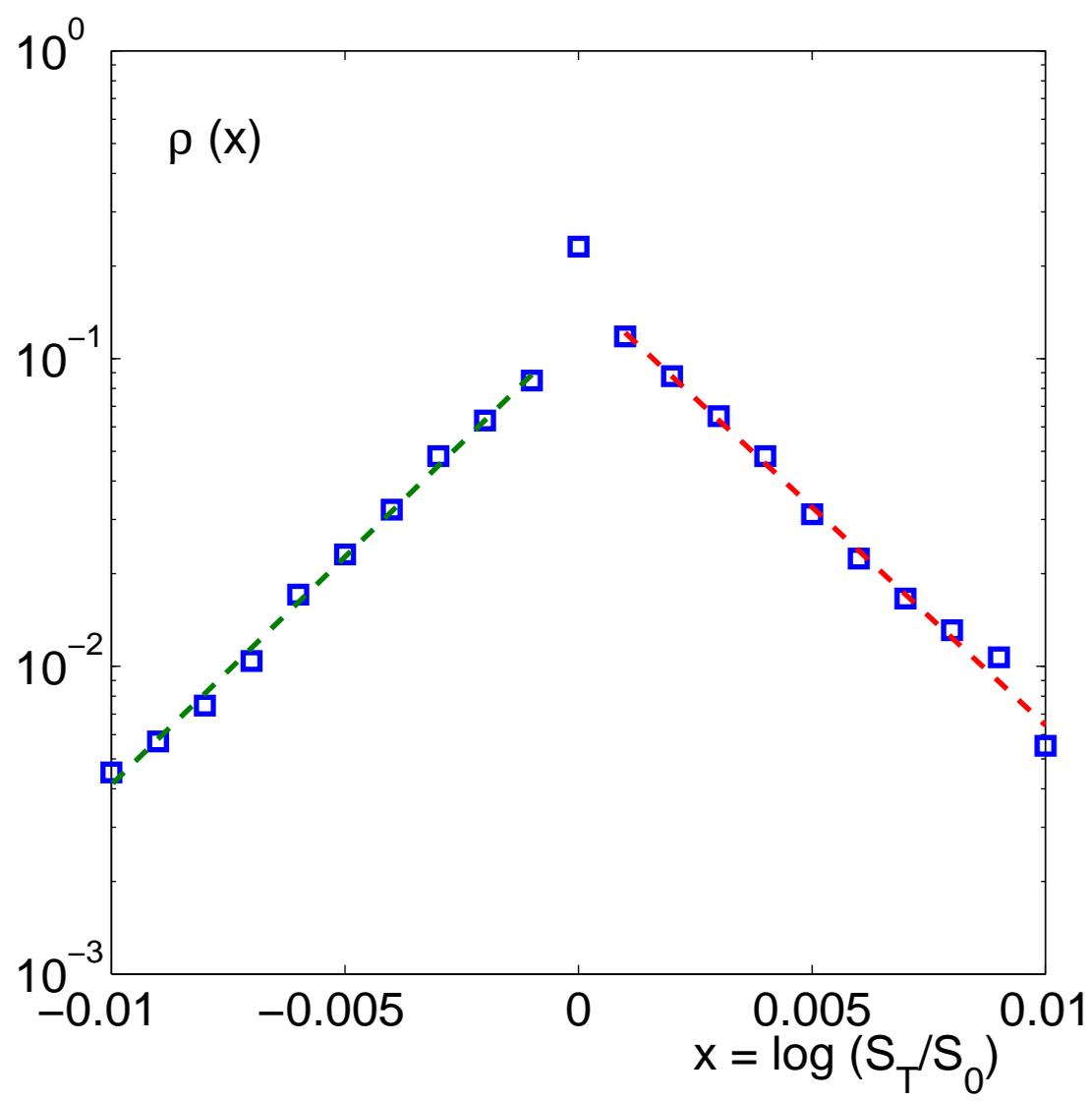

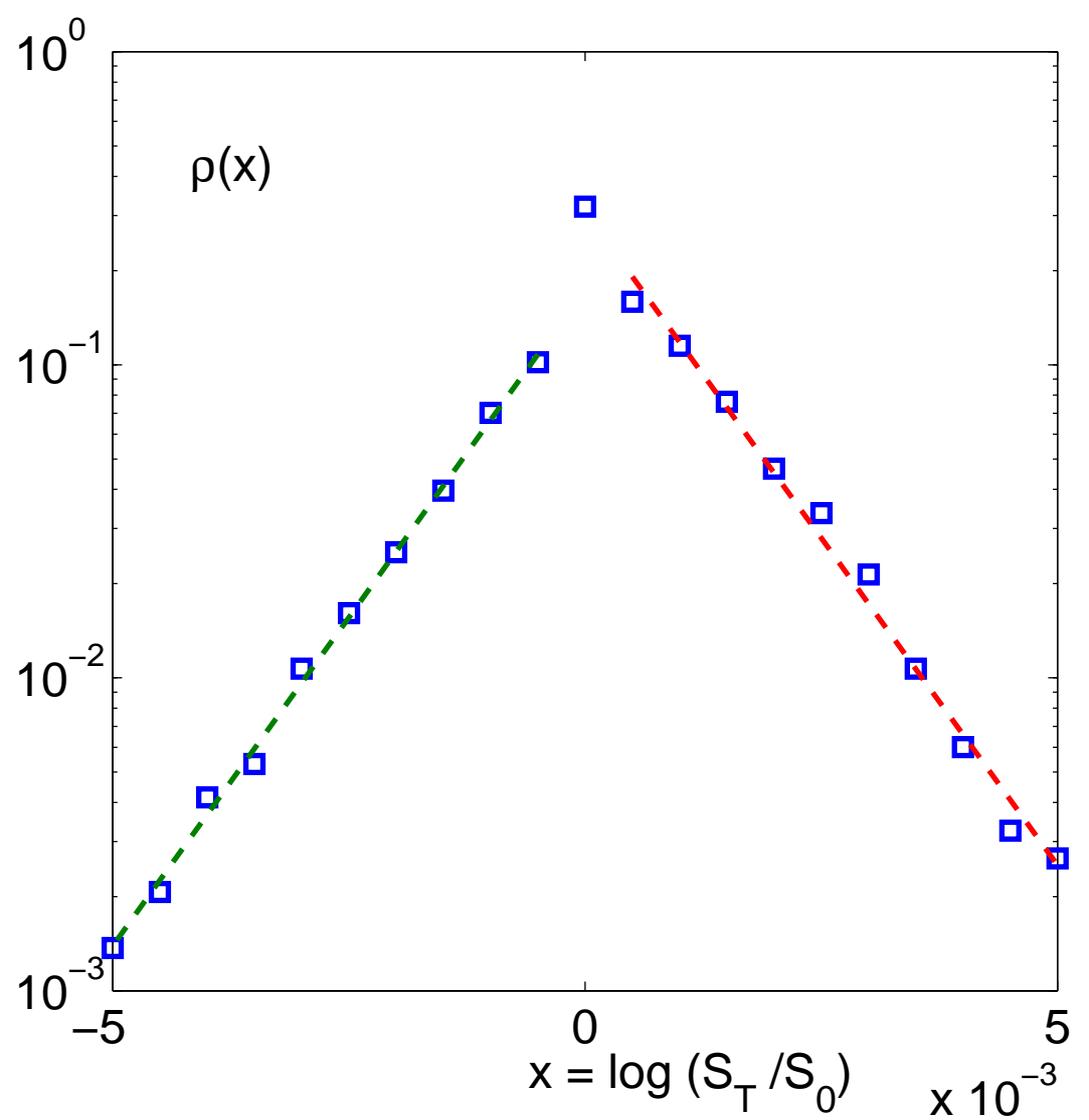

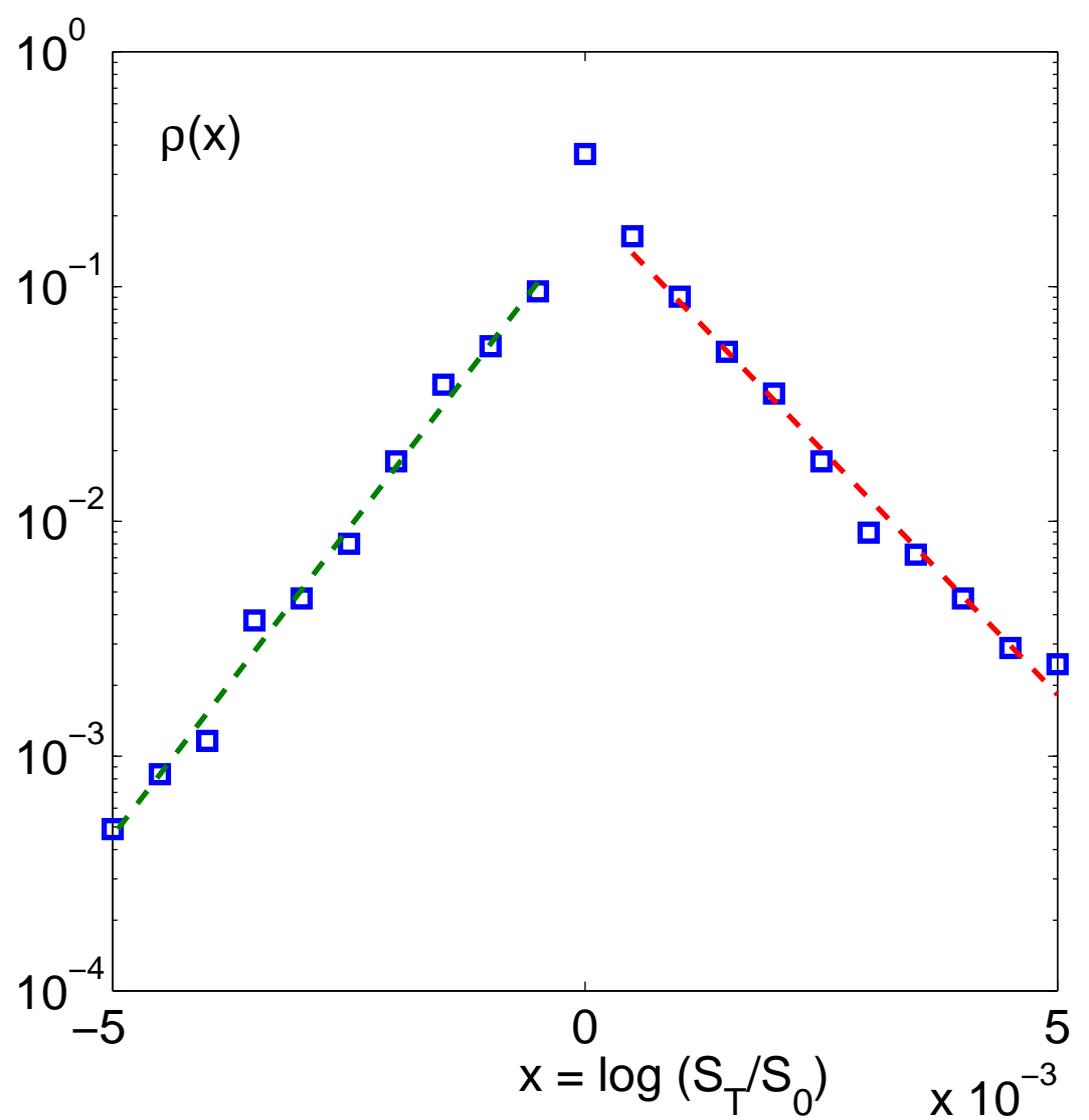

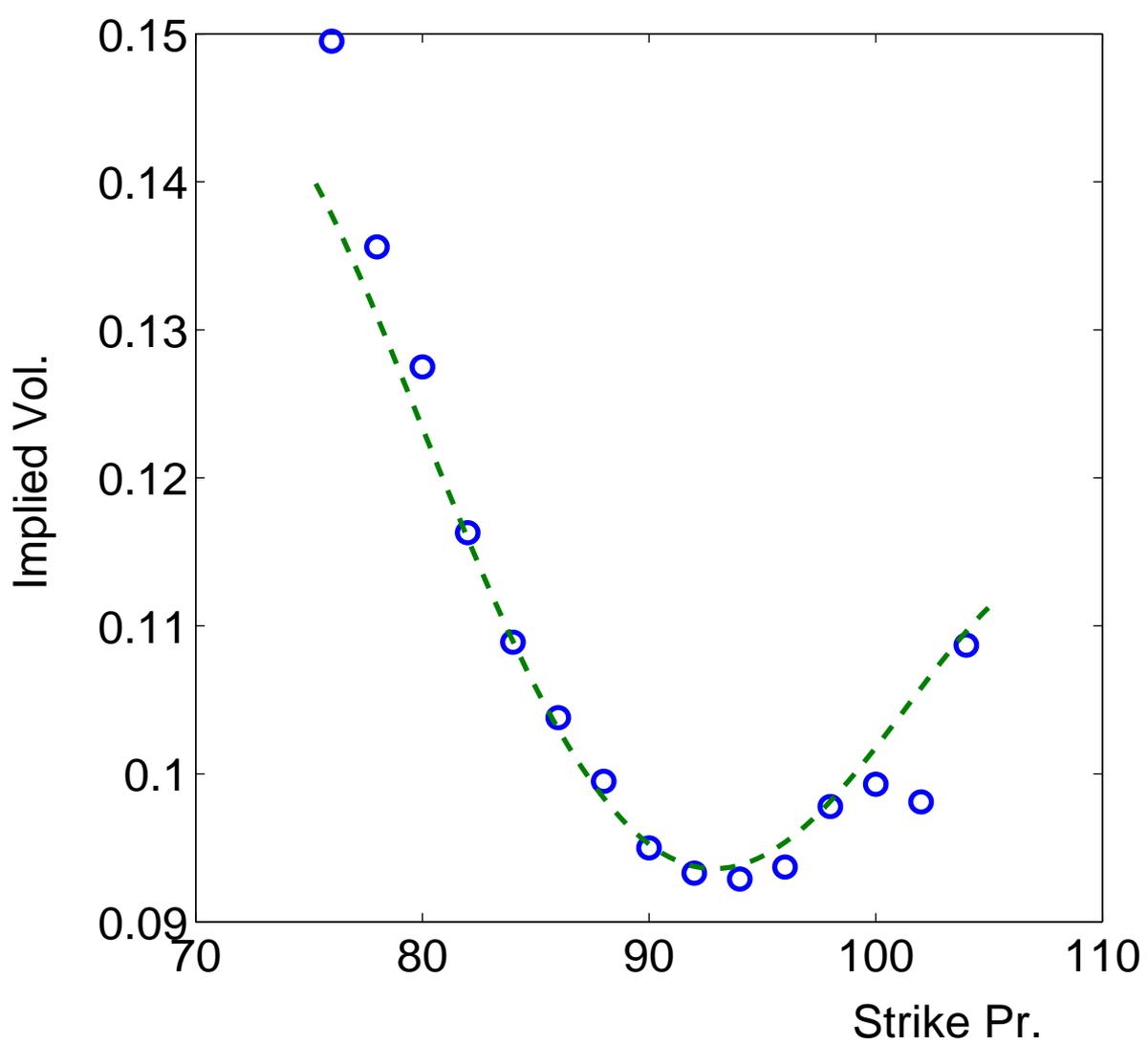